\documentclass[twocolumn,showpacs,preprintnumbers,amsmath,amssymb]{revtex4}

\usepackage{graphics}
\usepackage{epsfig}
\usepackage{bm}

\def\reff#1{(\ref{#1})}

\def\subsc#1{{\mbox{\rm\scriptsize #1}}}

\def\VKH{V_\subsc{KH}}

\def\vekt#1{\bm{#1}}

\def\vektr{\vekt{r}}

\def\vekte{\vekt{e}}
\def\vektE{\vekt{E}}

\def\vektA{\vekt{A}}

\def\vektalpha{\vekt{\alpha}}
\def\vektnabla{\vekt{\nabla}}

\def\halb{\frac{1}{2}}

\def\Edach{\hat{E}}
\def\Ehat{\Edach}
\def\Adach{\hat{A}}
\def\Ahat{\hat{A}}
\def\alphahat{\hat{\alpha}}

\def\energy{{\cal{E}}}

\def\pabl#1#2{\frac{\partial #1}{\partial #2}}
\def\bra#1{\langle #1 \vert}
\def\ket#1{| #1 \rangle}

\def\Veff{V^{\mbox{\rm\scriptsize eff}}}

\def\Atilde{\tilde{A}}

\def\imagi{\mbox{\rm i}}
\def\diff{\,\mbox{\rm d}}

\begin{document}

\title{Two-color stabilization of atomic hydrogen in circularly polarized laser fields}
\date{\today}
\author{D.~Bauer}
\email[E-mail: ]{dieter.bauer@physik.tu-darmstadt.de}
\affiliation{Theoretische Quantenphysik \& Theoretische Quantenelektronik, Institut f\"ur Angewandte Physik, Technische Universit\"at Darmstadt, Hochschulstr.\ 4a, 64289 Darmstadt, Germany}
\author{F.\ Ceccherini}
\affiliation{Dipartimento di Fisica ``Enrico Fermi'', Universit\`a di Pisa \& INFM, sezione A, Via Buonarroti 2, 56127 Pisa, Italy }
\date{\today}

\begin{abstract}
Dynamic stabilization of atomic hydrogen against ionization in high-frequency single- and two-color, circularly polarized laser pulses is observed by numerically solving the three-dimensional, time-dependent Schr\"odinger equation. The single-color case is revisited and numerically determined ionization rates are compared with both, exact and approximate high-frequency Floquet rates. The position of the peaks in the photoelectron spectra can be explained with the help of dressed initial states. 
In two-color laser fields of
opposite circular polarization the stabilized probability density may be shaped in various ways. For laser frequencies $\omega_1$ and  $\omega_2=n\omega_1$, $n=2,3,\ldots$ and sufficiently large excursion amplitudes $n+1$ distinct probability density peaks are observed. This may be viewed as the generalization  of the well-known ``dichotomy'' in linearly polarized laser fields, i.e, as ``trichotomy,'' ``quatrochotomy,'' ``pentachotomy'' etc.  All those observed structures and their ``hula-hoop''-like dynamics can be understood with the help of high-frequency Floquet theory and the two-color Kramers-Henneberger transformation. The shaping of the probability density in the stabilization regime can be realized without additional loss in the survival probability, as compared to the corresponding single-color results.
\end{abstract}


\maketitle

\section{Introduction}
For  sufficiently high laser frequency and intensity, the ionization probability of an atom decreases by further increasing the laser intensity. This, at first sight, counterintuitive effect is called dynamic or adiabatic stabilization, depending on the pulse length \cite{pont89,pontgav90,pontwalet,gavrevI,eberlykul,mittlemanbook,faisal,gavrilarevII}. So far, it has been observed in numerical simulations only (see \cite{kulander} and, for very recent results, \cite{dondera,choi}). Its theoretical explanation is most elucidating in the framework of high-frequency Floquet theory \cite{gavrevI}: the time-averaged Kramers-Henneberger potential, as it is seen by a free electron moving in the laser field, provides bound states. The electronic probability density of these states  is spread over spatial regions with dimensions of the order of the classical excursion amplitude which may be tens of atomic units or more. The laser frequency $\omega$, required for dynamic stabilization where the electron starts from the field-free ground state with binding energy $\energy_0$, has to exceed the field-dressed binding energy, $\hbar\omega>\vert\energy_0'\vert$. Since such high frequencies are not yet accessible to nowadays intense laser systems the experimental verification of dynamic stabilization is still to come. However, with the intense, short-wavelength, and coherent light sources under construction worldwide (e.g., at DESY in Hamburg) ground state stabilization should become experimentally accessible soon. Signatures of dynamic stabilization of Rydberg atoms, where $\vert\energy_0'\vert$ is smaller, have been measured \cite{druten}.

Dynamic stabilization has been studied extensively during the last decade. Ionization rates and energy shifts were obtained numerically by Floquet theory \cite{doerr} or approximate, analytical high-frequency Floquet theory \cite{pontgav90,gavrevI}. Full numerical {\em ab initio} solutions of the time-dependent, single-electron Schr\"odinger equation were obtained for low-dimensional model systems (see, e.g., \cite{su,volkova,patelI,patelII,kwon,ryab}) and, after the early work in \cite{kulander} very recently, in three spatial dimensions for linear polarization \cite{dondera} and circularly polarized laser pulses \cite{choi}. Although the Floquet ionization {\em rates} decrease monotonically for laser intensity $I\to\infty$ the ionization {\em probability} in full {\em ab initio} solutions of the TDSE for {\em finite} laser pulses assumes a minimum and increases again thereafter. In the limit of laser intensity $I\to\infty$ but otherwise  fixed pulse parameters the survival probability of an atom will thus approach zero.   Moreover, it has been found that the magnetic field of the laser---usually neglected because of the dipole approximation---leads to increased ionization \cite{kylstra}. This is due to the ponderomotive force which pushes the electrons in propagation direction.  In multi-electron atoms electron correlation also hinders stabilization (see \cite{bauer}, and references therein). Two-color interference effects on stabilization in linear polarization was studied in \cite{cheng,potvl}. 
   
In this paper the dynamic stabilization of H in circularly polarized single- and two-color fields is investigated. The article is organized as follows: in Sec.~\ref{theory} the three-dimensional, time-dependent Schr\"odinger equation to be solved numerically is introduced, and the form of the applied two-color fields is explained. In Sec.~\ref{singlecolor} the results concerning the single-color, circular polarization  case are presented. Besides snapshots of the probability density and photoelectron spectra, particular emphasis is put on the comparison of ionization rates and energy shifts with results from Floquet theory. In Sec.~\ref{twocolor} the main two-color results are presented. Probability density structures are observed which display well separated maxima, the number of which is related to the frequency ratio of the two fields.  The structure  moves as a whole in the laser field and can be explained with the help of two-color, time-averaged Kramers-Henneberger potentials. Survival probabilities after the two-color pulses are presented as a function of the two laser's peak field strengths. The two-color survival probabilities and above-threshold ionization spectra are compared with the corresponding single-color results where the second field was absent. It is good news that adding a second color is possible without a drastic inhibition of stabilization even if the second laser's intensity lies in the ``death valley'' of high ionization. Finally, we conclude in Sec.~\ref{concl}.

\section{Theory} \label{theory}
The time-dependent Schr\"odinger equation (TDSE) for the electron of
atomic hydrogen in the electric field of a laser
\begin{equation} \imagi \pabl{}{t} \Psi(\vektr,t) = \left( \halb [
-\imagi\vektnabla + \vektA(t)]^2  -\frac{1}{r} \right)   \Psi(\vektr,t) \label{tdse}\end{equation}
is solved numerically in three spatial dimensions (atomic units are
used unless noted otherwise). The intense laser field can be treated classically. Since the laser wavelength is large compared to all other relevant length scales and the electron dynamics is non-relativistic for our laser parameters, the dipole
approximation can be applied so that the vector potential $\vektA(t)$ does
not depend on spatial coordinates. Hence, the magnetic component of the laser field is neglected. 
If we restrict ourselves to study {\em coplanar} two-color
fields (of, however, arbitrary ellipticity) we may choose the coordinate system in such a way that the  $z$-component of the electric field $\vektE(t)=-\partial_t\vektA(t)$ vanishes: $\vektE(t)=\vektE^{(1)}(t)+\vektE^{(2)}(t)=[E_x^{(1)}(t)+E_x^{(2)}(t)]\vekte_x+[E_y^{(1)}(t)+E_y^{(2)}(t)]\vekte_y$.  

The wave function $\Psi(\vektr,t)$ is expanded in spherical harmonics, 
 \[ \Psi(r,\vartheta,\varphi,t)=\frac{1}{r} \sum_{\ell=0}^\infty
\sum_{m=-\ell}^\ell {\Phi_{\ell m} (r,t)} Y_{\ell m} (\vartheta,\varphi),
\]
which leads to a set of coupled TDSEs for the radial wave functions $\Phi_{\ell m}(r,t)$,
\begin{widetext}
\begin{eqnarray} \quad \hspace{-1cm} \imagi \pabl{}{t} {\Phi_{\ell m}}(r,t) &=& \left( -\halb {\pabl{^2}{r^2}} +
\Veff_\ell(r) \right)  \Phi_{\ell m}(r,t) \nonumber \\
&& +\imagi \sqrt{\frac{2\pi}{3}} \sum_{\ell' m'} \Biggl\{ {\Atilde^*(t)} \langle
  \ell m | 1 1 | \ell' m' \rangle \pabl{}{r} - {\Atilde(t)} \langle \ell m | 1 -1 |
  \ell' m' \rangle \pabl{}{r} \label{tdseII} \\
&& \qquad  -\frac{{\Atilde^*(t)}}{r} \langle \ell m | 1 1 | \ell' m'
  \rangle (1+m') + \frac{{\Atilde(t)}}{r} \langle \ell m | 1 -1 | \ell' m'
  \rangle (1-m') \nonumber \\
&&  \qquad  -\frac{{\Atilde^*(t)}}{r} \langle \ell m | 1 0 | \ell' m'+1
  \rangle {N_{\ell' m'}^+} + \frac{{\Atilde(t)}}{r} \langle \ell m | 1 0 | \ell' m'-1
  \rangle {N_{\ell' m'}^-} \Biggr\} {\Phi_{\ell' m'}}(r,t)\nonumber 
\end{eqnarray}
\end{widetext}
where
$ \Veff_\ell(r)=-\frac{1}{r} + \frac{\ell(\ell+1)}{2 r^2} $
is the effective atomic potential including the centrifugal barrier,
$\Atilde(t)= A_x(t)+\imagi A_y(t)$, ${N_{\ell m}^\pm} = \sqrt{(\ell\mp m)(\ell\pm m
      +1)/2}$, and the purely time-dependent $\vektA^2$-term has been transformed away.
The matrix elements $\bra{\ell m} LM \ket{\ell'
m'}:=\int\diff (\cos\vartheta) \diff\varphi\, Y^*_{\ell m}(\vartheta,\varphi) Y_{L M}
(\vartheta,\varphi) Y_{\ell' m'} (\vartheta,\varphi)$ may be expressed
in terms of Clebsch-Gordan coefficients.


It is worth noticing the coupling of different $\ell$ and $m$ quantum
numbers in the TDSE \reff{tdseII} due to the laser field: there is a
$\ell m \to \ell\pm 1, m\pm 1$ coupling as well as a $\ell m \to
\ell\mp 1, m\pm 1$ coupling (where compatible with the condition
$-\ell\leq m \leq \ell$). From the knowledge of these couplings one
can immediately deduce non-perturbative selection rules, e.g., if the initial wave function is the 1s groundstate, states with ($\ell=1$, $m=0$), with ($\ell=2$, $m=\pm 1$), with ($\ell=3$, $m=\pm 2$) etc.\ are `unreachable' and thus remain unoccupied. 

In the actual implementation, the radial coordinate $r$ was
discretized directly in position space. The details of the
implementation may be found in \cite{bauerhabil} and is similar to what has been
described in \cite{mullermethod} for the simpler case of linear polarization.

The radius of our numerical grid, including the region where the non-vanishing imaginary potential damps away probability density approaching the grid boundary, was between 240\,a.u.\ and 450\,a.u. The maximum $\ell$ quantum number was typically $\ell_{\max}=30$. For testing the convergence of our results runs up to $\ell_{\max}=50$ were also performed---with no significant changes in the observables of interest.  
For obtaining the results presented in this paper, the run times of
our Schr\"odinger-solver were always less than 18 hours on 1 GHz PCs.

We focus on two-color fields where both fields are
circularly polarized. We assume a $\sin^2$-shaped up- and down-ramp of
the $i$th field, $i=1,2$ over $N_1^{(i)}=N^{(i)}_{23}$ cycles and a
constant part of $N^{(i)}_{12}$ cycles duration. In the following we
will refer to such a pulse as a
$(N_1^{(i)},N^{(i)}_{12},N^{(i)}_{23})$-pulse.
For the $i$th electric field in, say, $x$-direction  
\begin{equation}  E_x^{(i)}(t)= \left\{ \begin{array}{lcl}
    E_\subsc{ud}^{(i)}(t) & \mbox{\rm for} & 0 \leq t < T_1, \\
    \Ehat^{(i)} \cos \omega_i t & \mbox{\rm for} &  T_1 \leq t < T_2, \\
    E_\subsc{ud}^{(i)}(t-T_2+T_1) & \mbox{\rm for} & T_2 \leq t < T_3, \\
0 & \mbox{\rm otherwise}  & \\
\end{array} \right.\label{xcomponent} \end{equation}
is set where $E_\subsc{ud}^{(i)}(t) = \Ehat^{(i)}\sin^2(\pi t/ 2 T_1) \cos \omega_i t$. The absolute ramping time as well as the duration of the constant part of the laser pulse are chosen the same for both colors so that the times $T_1=2 \pi N_1^{(i)}/\omega_i$, $T_2=T_1+2\pi N_{12}^{(i)}/\omega_i$, etc.\ are independent of $i$. 
Since we are interested in {\em opposite} circular polarizations 
\begin{equation} E_y^{(1)}(t)= E_x^{(1)}(t-3\pi/2\omega_1),\ E_y^{(2)}(t)= E_x^{(2)}(t-\pi/2\omega_2) \label{ycomponents} \end{equation}
are taken as the $y$-components of the two colors.
The corresponding vector potential can be calculated from $\vektA(t)=-\int_0^t\diff t'\, \vektE(t')$. In such two-color pulses, as long as
$N_1^{(i)}=N^{(i)}_{23}$, $N_{12}^{(i)}$ are integer numbers,
a {\em free} electron (initially at rest)  returns to its
starting point at the end of the laser pulse---without residual drift momentum.

\section{The circular single-color case revisited} \label{singlecolor}
Before turning to the two-color study let us analyze the simpler single-color case. For the latter, results from {\em ab initio} solutions of the three-dimensional TDSE were obtained only recently  \cite{dondera} whereas two-dimensional model solutions were presented earlier \cite{patelI,patelII,kwon,ryab}.
Ionization {\em rates} as well as energy shifts for
hydrogen in the stabilization regime (with dipole approximation) are
available for several years \cite{doerr} so that a comparison with our finite
pulse results is possible and will be presented in this paper. Since all vector potential amplitudes $\Ahat$ considered in this work are $\leq 10$\,a.u.\ the dipole approximation is applicable, as has been shown in \cite{kylstra}.

\subsection{Survival probability vs.\ time}
In order to calculate the probability for the hydrogen atom to ``survive'' non-ionized till the end of the pulse the probability density is integrated over a spherical volume of radius $R_b=60$ atomic units. The radius $R_b$ was chosen several times greater than the biggest excursion amplitude $\hat{\alpha}=10$\,a.u.\ considered in this paper. Since the free part of the wavefunction vanishes inside that sphere at sufficiently large times after the laser pulse has passed by, the amount of probability density inside the sphere leads then with high accuracy to the same probability one would obtain by projecting on all bound states.   
\begin{figure}
\includegraphics[scale=0.75]{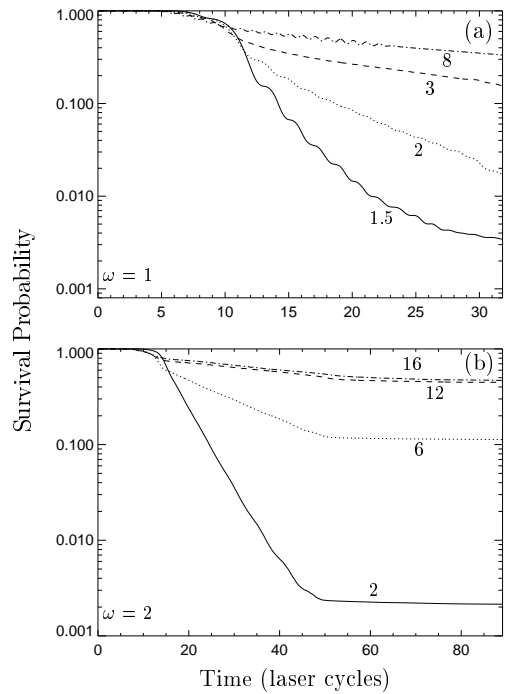}
\caption{\label{severalA:bigsphere} The probability density, integrated over  a sphere of radius $R_b=60$  for laser pulses of frequency $\omega=1$  [plot (a)] and $\omega=2$ [plot (b)] vs.\ time. (4,16,4)- and (4,32,4)-pulses were chosen for $\omega=1$ and $\omega=2$, respectively. The $\Ehat$ values are attached to the curves. The survival probability increases with increasing peak field strengths, but not beyond a certain maximum. In some of the cases an ionization  rate can be determined from the (in the logarithmic plot) linear slope. }
\end{figure}
The results are presented in Fig.~\ref{severalA:bigsphere}. In the upper panel (a) the results for $\omega=1$ are presented, in the lower panel (b) those for  $\omega=2$. All pulses were up- and down-ramped over 4 cycles. The constant part of the pulse lasted 16 cycles for $\omega=1$ and 32 cycles for $\omega=2$. By Fig.~\ref{severalA:bigsphere} the very existence of dynamic stabilization is proven:  the survival probability at the end of the pulse increases with increasing peak field strength (attached to the curves) up to a maximum value. Increasing the peak field strength further yields a decreasing survival probability (not shown in Fig.~\ref{severalA:bigsphere}). The highest survival probability in plot (b) amounts to $\approx 47\%$. It should be noted, however, that the survival probability is rather sensitive to the laser pulse shape and duration. This was also found  in the case of linear polarization and is discussed at length in Ref.~\cite{dondera}.

Whenever the slope of the probability curves vs.\ time is linear on the logarithmic scale an ionization {\em rate} can be determined. While this is straightforward for the curves in the $\omega=2$ panel it turns out to be more difficult for the $\omega=1$-results in (a). Longer pulses of constant intensity are then needed to obtain unambiguous rates.

\subsection{Snapshots of the probability density}
For sufficiently large excursions
$\hat{\alpha}$ and adiabatic turn-ons one expects the system to occupy
essentially the ground state in the time-averaged Kramers-Henneberger
(KH) potential. This corresponds to lowest-order high-frequency
Floquet theory \cite{gavrevI}. The KH potential is the ionic potential a free
electron, just oscillating in the laser field, would ``see'',  i.e.,
$\VKH(\vektr)=V(\vektr + \vektalpha(t))$ where $\vektalpha(t)$
is the trajectory of a free electron in the field 
\begin{equation}\vektalpha(t)=\int_0^t\diff t'\, \vektA(t').\end{equation} 
In the case of a single-color, circularly polarized laser the KH
potential, averaged over one laser cycle, has azimuthal symmetry with respect to the laser propagation direction (the $z$-axis in our case) and,
for sufficiently big $\hat{\alpha}$, its minimum along a circle of
approximate radius $\hat{\alpha}$ in a plane perpendicular to it (the $xy$-plane in our case). Consequently, the KH ground state probability density has
its maximum along that circle. By transforming back to the lab frame, where the ionic potential is assumed centered at the origin, 
this ring-shaped, stabilized probability density 
orbits with its center along the free electron trajectory which itself
is a circle of radius $\hat{\alpha}$. What results is a
``hula-hoop''-like dynamics \cite{choi} of a probability density ring about the
ion which assumes the role of the ``hula-hoop''-dancer's hip.

\begin{figure}
\includegraphics[scale=0.5]{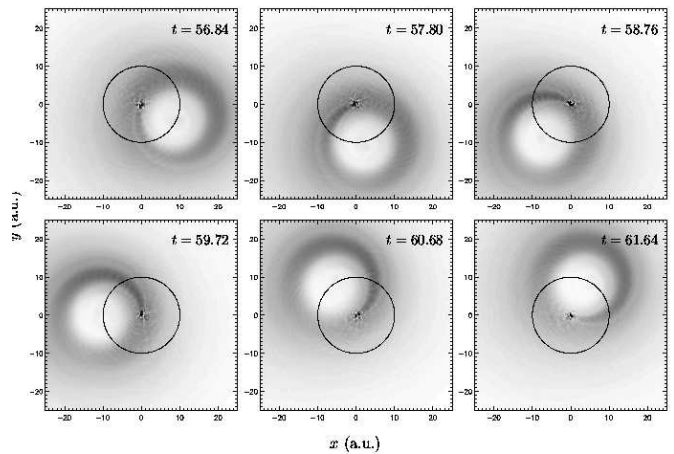}
\caption{\label{snapshots:onlyw1:10} Series of contour plots
of the probability density in the $xy$-plane at $z=0$ during the 9th cycle of the (2,8,2)-laser pulse for $\omega_1=1$ and
$\Edach_1=10$. The color-coding of the probability density is linear. The overlayed black circle of radius $\hat{\alpha}=10$ indicates the corresponding free electron trajectory. A clockwise ``hula-hoop''-like \cite{choi} dynamics of the probability density ring is observed.  }
\end{figure}

In Fig.~\ref{snapshots:onlyw1:10} we present snaphots of the
probability density in the $xy$-plane at $z=0$ during the 9th laser
cycle of the (2,8,2)-pulse with $\omega=1$. The peak field strength was $\Ehat=10$, yielding  an excursion radius $\hat{\alpha}=10$. Although a probability density ring, describing the previously explained ``hula-hoop''-dynamics, is clearly visible, the probability density is not uniformly distributed along that ring but has rather a sickle-like shape. Hence, the system is obviously not in the KH ground state but in a superposition of states with different azimuthal quantum numbers. This leads to Rabi floppings on time scales that are large compared to the laser period. Note that  there is always a probability density maximum at the position of the ion.
For lower values of $\alphahat$ we observed more complex situations  where no clear ring structure but two or more wave packets swirl about the ion.

\subsection{Ionization rates} \label{ionirates_singlecolor}
For the single-color case, accurate ionization rates were numerically
obtained many years ago by means of non-perturbative Floquet theory
\cite{doerr}. Those rates are accurate in the sense that {\em if} the system
somehow manages to occupy a single Floquet state it will decay
precisely with the corresponding Floquet rate. Usually an adiabatic
turn-on is needed to transfer the system from the field-free
groundstate to a single Floquet state. The problem is
that an adiabatic ramping of the laser pulse counteracts stabilization
because the system has to spend too much time in the so-called ``death
valley'' of medium laser intensities where ionization is strong, and
thus nothing would survive to be stabilized. Hence, short turn-on times are
desirable. However, after such rapid turn-ons the system might be ``shaken-up'' into a
superposition of many Floquet states. Under such conditions only the full solution of the TDSE yields reliable ionization probabilities and, if existing, ionization rates. 

\begin{figure}
\includegraphics[scale=0.5]{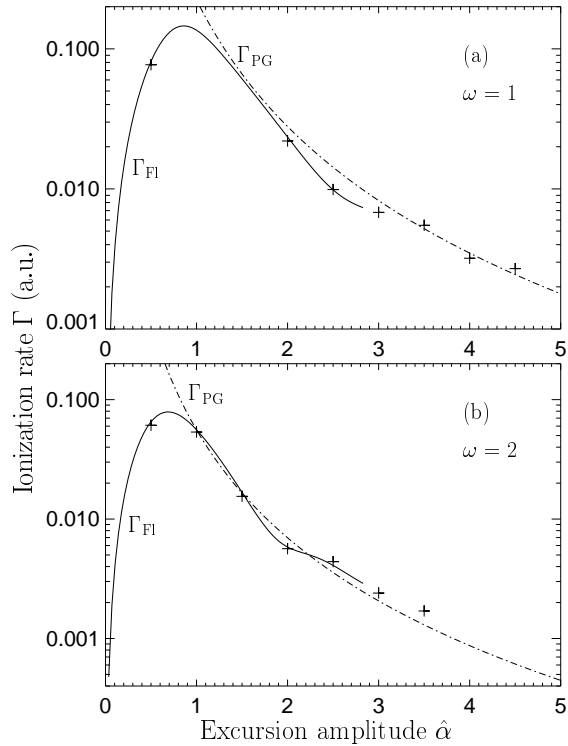}
\caption{\label{ratescomp} Ionization rates vs.\ excursion amplitude $\hat{\alpha}$ for (a) $\omega=1$ and (b) $\omega=2$\,a.u. Our numerical results (+), the full Floquet rates $\Gamma_{\subsc{Fl}}$ from \cite{doerr} (solid), and the analytical result by Pont and Gavrila \cite{pontgav90,gavrevI} $\Gamma_{\subsc{PG}}$ (dashed-dotted) are shown.  }
\end{figure}

In Fig.~\ref{ratescomp} our ionization rates for $\omega=1$ and
$\omega=2$ (determined from plots like those in
Fig.~\ref{severalA:bigsphere}) are compared with Floquet
results \cite{doerr} and the analytical results derived by Pont and Gavrila \cite{pontgav90,gavrevI}. The Pont-Gavrila rate for circular polarization 
\begin{equation} \Gamma_{\subsc{PG}}\approx\frac{0.223}{\hat{\alpha}^4 \omega^2} \end{equation} 
was derived from high-frequency Floquet theory combined with the  first-order Born approximation. In order to meet the plane wave criterion of  the Born approximation the excursion amplitude $\hat{\alpha}$ must not be too small. For high frequencies and $\hat{\alpha}\gg 1$ the Pont-Gavrila rate should merge with the full Floquet result. However, in \cite{doerr} Floquet rates were calculated only for moderate values of $\hat{\alpha}$ [note, that, by definition, the excursion amplitude $\hat{\alpha}$ for circular polarization in \cite{doerr} differs by a factor $\sqrt{2}$ with Pont's and Gavrila's, and ours, i.e., $\Gamma_{\subsc{PG}}= \Gamma_{\subsc{Fl}}/\sqrt{2}$]. For $\omega=2$ the Pont-Gavrila rate comes close to the full Floquet result already for $\hat{\alpha}=1$. For increasing $\hat{\alpha}$ the full Floquet result oscillates around the Gavrila rate.  In the lower frequency case $\omega=1$ the Gavrila rate overestimates ionization. Our numerical rates were obtained from simulations of (4,16,4)-pulses for $\omega=1$ and (4,32,4)-pulses for $\omega=2$. The results for the highest $\hat{\alpha}$ values shown were difficult to determine because the ionization probability curves vs.\ time displayed no clear exponential decrease. In order to determine unambiguous ionization rates also for higher $\hat{\alpha}$ a more adiabatic up-ramp of the laser field and a longer pulse duration would be required. However, since it is not the goal of this paper to determine rates which are probably easier to obtain via Floquet theory, we do not proceed further in this direction. Our numerical results agree well with the full Floquet rates, where available. For $\omega=1$ and $ \hat{\alpha}\geq 3.5$ the Gavrila rate agrees quite well with our numerical data whereas for $\omega=2$ it underestimates ionization. 

\subsection{Above-threshold ionization (ATI)} \label{ati_singlecolor}
\begin{figure}
\includegraphics[scale=0.53]{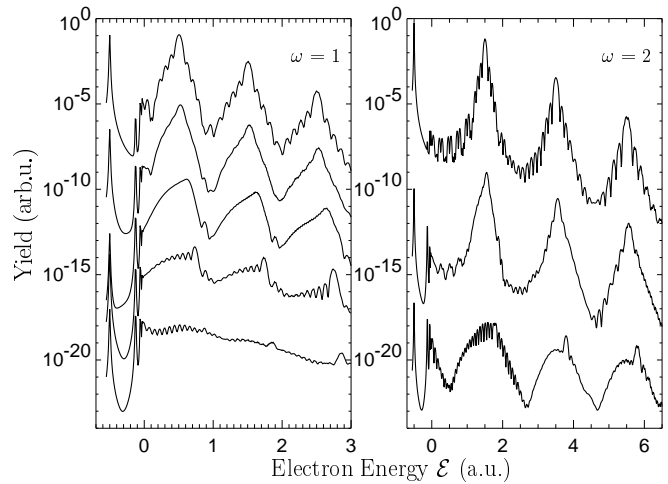}
\caption{\label{atispectra} ATI photoelectron spectra for $\omega=1$
(left panel) and $\omega=2$ (right panel) after (4,8,4)-pulses (for
$\omega=1$) and (8,16,8)-pulses (for $\omega=2$). For clarity, the spectra are shifted vertically with respect to each other. The peak field
strengths were (left panel, from top to bottom) $\Edach=0.3,0.5,1,2,5$
and (right panel, from top to bottom) $\Edach=0.6,2,10$ atomic units.
For certain laser intensities excited states come into play, leading
to pronounced modulations of the ATI peaks. With increasing laser intensity the ATI peaks move towards higher energies.} 
\end{figure}

In Fig.~\ref{atispectra} photoelectron spectra after
(4,8,4)-pulses (for $\omega=1$) and (8,16,8)-pulses
 (for $\omega=2$) of different peak field strength are presented. They were calculated from the wave function immediately after the pulse with the
window-operator method proposed in \cite{windowop}. The three lowest-order ATI peaks $n=1,2,3$, separated by $\hbar\omega$, are clearly visible although they  broaden with increasing laser intensity. 
From lowest order perturbation theory (LOPT) an exponential decrease of the peak heights $\sim \Edach^{2n}$ with increasing $n$ is expected. Within high-frequency Floquet theory, for $\hat{\alpha}\gg1$, $(\energy_0'+n\omega)\hat{\alpha}\gg1$ the contributions of higher orders $n$ do not vanish exponentially but only $\sim n^{-2}$ \cite{pontgav90}. By virtue of the first two peaks in the highest-$\Ehat$ spectra of Fig.~\ref{atispectra} this trend of increasing importance of higher order peaks is confirmed [note that in the numerically obtained wavefunction after the pulse the ATI peak $n=3$ might be slightly suppressed because of the absorbing boundaries]. 
With increasing population
of excited states the left shoulder of the ATI peaks becomes more
pronounced. For certain laser intensities a substructure of smaller
peaks appears. Since these peaks are separated by the energy
difference of $n=2$ and $n=3$-states, that is $\Delta \energy
=(4^{-1}-9^{-1})/2\approx 0.07$, this substructure is likely connected
to the population of these states. However, we did not perform yet a detailed study of this
substructure. Here we want to focus on the ATI peak {\em positions}. As can
be easily inferred from Fig.~\ref{atispectra} the ATI peaks move with
increasing  laser intensity to higher photoelectron energies. This is
contrary to what happens for low frequencies and high intensities.
In  general, the positions of the ATI peaks follow   
\begin{equation} \energy_\subsc{low}^{(n)}=\energy_0' - \Delta_\subsc{cont} + n \omega .\label{tunnelpeakpos}\end{equation} 
Here, $\energy_0'=\energy_0+\Delta$ is the field-dressed ground state
energy which is ac Stark-shifted by $\Delta$ compared to the unperturbed ground
state energy $\energy_0$ ($=-0.5$ for H), and $\Delta_\subsc{cont}$ is the energy up-shift of the continuum threshold. $n$ is 
the number of absorbed photons $\geq n_{\min}$ where $n_{\min}$ is the smallest integer that
gives $\energy_\subsc{low}^{(n_{\min})}>0$. For low frequencies one has $\Delta_\subsc{cont}=U_p=f \frac{\Ehat^2}{2\omega^2}$ where $U_p$ is the ponderomotive
potential, i.e., the time-averaged quiver energy of a free electron in
the laser pulse of peak field strength $\Edach$. The pre-factor $f$ is (for our definition of
$\Edach$) $1$ for circular and $1/2$ for linear polarization. Since in the case of long wavelengths the
up-shift of the continuum threshold by $U_p$ is bigger than the shift
$\Delta$ of the ground state, the  ATI peaks move towards lower energies
with increasing laser intensities. The ``drowning'' of a certain peak
no.\ $n$ below $\energy=0$ is the celebrated ``channel closing'' in ATI
physics.

We shall see that in the stabilization regime, where $\omega > \vert\energy_0'\vert$ must
hold, the shift of ATI peaks is dominated by the ground state up-shift $\Delta$ so that 
\begin{equation}  \energy_\subsc{high}^{(n)} \approx \energy_0' + n\omega . \label{peakposanalyt}\end{equation}
Accurate values of the field dressed energies $\energy_0'$ in the stabilization regime can be obtained from Floquet theory \cite{doerr}. 
Solving the set of coupled
Floquet equations yields complex energies the real part of which is
the field-dressed energy. The complex part is half the
ionization rate,
$\energy_{0,\subsc{Fl}}=\energy_0'-\imagi\Gamma_\subsc{Fl}/2$.
From the high-frequency Floquet theory point-of-view  $\energy_0'$ is the groundstate
energy of the time-averaged KH potential. In Ref.~\cite{pontgav90} Pont and Gavrila gave KH
ground state energies for H in a circularly polarized laser field.
Because $\vert\energy_0'\vert < \vert\energy_0\vert$ the ATI peaks move to higher energies with increasing laser intensity---contrary to the high-intensity, low-frequency-case \reff{tunnelpeakpos}.

\begin{figure}
\includegraphics[scale=0.5]{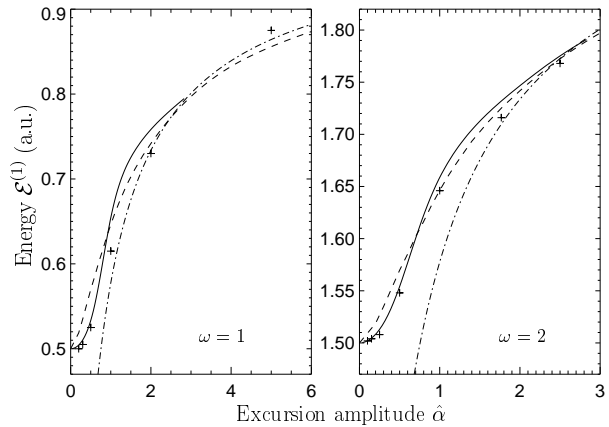}
\caption{\label{peakposfig} The energy $\energy^{(1)}$ of ATI peak
$n=1$ vs.\ the excursion amplitude $\hat{\alpha}=\Edach/\omega^2$ for
$\omega=1$ (left plot) and $\omega=2$ (right plot). The + symbols are
our numerical results. The curves give the expected position of ATI peak $n=1$  according
\reff{peakposanalyt} with $\energy_0'$ from the full Floquet
calculation \cite{doerr} (solid line), with $\energy_0'$ as the ground state
of the time-averaged KH potential \cite{pontgav90} (dashed), and with
$\energy_0'=-(\ln\hat{\alpha} + 2.654284)/(2\pi\hat{\alpha})$ \cite{pont89} (dashed-dotted). }
\end{figure}

In Fig.~\ref{peakposfig} our numerically determined positions
for the $n=1$ ATI peak 
are compared with Eq.~\ref{peakposanalyt} where the value for
$\energy_0'$ has been determined   (i) by the full Floquet solution
\cite{doerr}, (ii) as the time-averaged KH potential ground state energy \cite{pontgav90},
and  (iii) through Pont's approximate formula
$\energy_0'=-(\ln\hat{\alpha} + 2.654284)/(2\pi\hat{\alpha})$ \cite{pont89}.
The agreement may be considered  acceptable although not excellent. 
The differences are likely due to the population of more than a single Floquet state.
Our numerical result for $\omega=1$ and $\hat{\alpha}=5$ lies above the analytical curves because the substructure in the ATI spectrum (cf.\ Fig.~\ref{atispectra}) also alters the absolute maximum of an ATI peak so that its determination becomes ambiguous.

\section{Two-color case with opposite circular polarizations} \label{twocolor}
\subsection{Time-averaged KH potentials}
\begin{figure}
\includegraphics[scale=0.45]{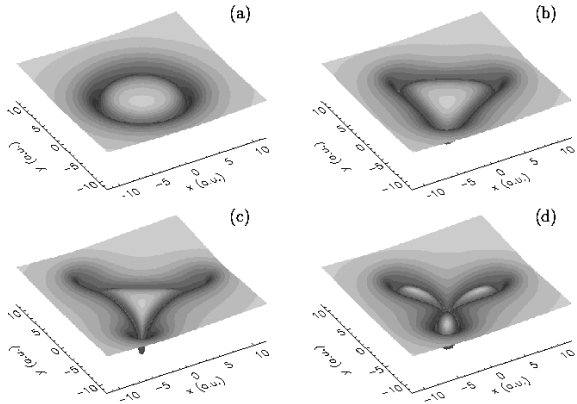}
\caption{\label{timeavKHpots} Time-averaged KH potentials in the $xy$-plane at $z=0$ for $\omega_1=1$, $\omega_2=2$. The peak field strengths
$\Ehat^{(1)}$, $\Ehat^{(2)}$ are chosen in such a way that (a) the
$\omega_2$-field is absent, $\Ehat^{(1)}=6$; (b) the two laser
intensities are equal, $\Ehat^{(1)}=\Ehat^{(2)}=6$; (c) the two
vector potential amplitudes are equal, $\Ehat^{(1)}/\omega_1=\Ehat^{(2)}/\omega_2=6$;
(d) the two excursion amplitudes are equal,
$\Ehat^{(1)}/\omega_1^2=\Ehat^{(2)}/\omega_2^2=4$. 
In the
two-color cases (b)--(d) the three-fold symmetry leads to three minima
in the potential. The corresponding KH ground state probability
densities will therefore display ``trichotomy,'' i.e., three spatially
separated maxima of the probability density.}
\end{figure}

It is useful for the analysis of our two-color results to get familiar
with the free electron trajectories in such fields and the corresponding
time-averaged KH potentials. In the two-color case the KH potentials were averaged over the low-frequency laser period $2\pi/\omega_1$. Let us consider the two laser frequencies $\omega_1=1$ and $\omega_2=2$. If the second field is absent the free electron trajectory $\vektalpha(t)$ describes a circle in the $xy$-plane of radius $\hat{\alpha}_0=\Ehat^{(1)}/\omega_1^2$. If the second field amplitude $\Ehat^{(2)}$ equals $\Ehat^{(1)}$, i.e., both laser fields have the same intensity, the trajectory resembles an equilateral triangle with rounded edges, and $\max\vert\hat{\alpha}\vert=5\hat{\alpha}_0/4$. If, instead, the vector potential amplitudes of both fields are equal, $\Ehat^{(1)}/\omega_1=\Ehat^{(2)}/\omega_2=$, the electron follows a convex triangle with sharp edges, and $\max\vert\hat{\alpha}\vert=3\hat{\alpha}_0/2$. Finally, equal excursion amplitudes, i.e., $\Ehat^{(1)}/\omega_1^2=\Ehat^{(2)}/\omega_2^2$, yields a rosette-like trajectory, and $\max\vert\hat{\alpha}\vert=2\hat{\alpha}_0$. The time-averaged KH potentials look accordingly. They are presented in Fig.~\ref{timeavKHpots}. In the two-color cases (b)--(d) the
three-fold symmetry leads to three potential minima near the turning
points. If the stabilized electron occupies the KH ground state the
probability density will show ``trichotomy'' instead of the well-known
``dichotomy'' in the case of a single-color linearly polarized laser
where two probability peaks, separated by $2\hat{\alpha}$ were observed \cite{kulander}. In general, when $\omega_2=n\omega_1$ the trajectories display $(n+1)$-fold symmetry. 
Provided the peak field strengths are properly chosen, for $n=3,4,\ldots$ the $n+1$ minima of the time-averaged KH potentials will lead to ``quatrochotomy,'' ``pentachotomy,'' etc. The electron dynamics in the lab frame should then resemble ``hula-hoop'' with a triangle, a square, a pentagon, and so forth.  
We would like to remark that the discrete symmetry of the trajectory and the time-averaged KH potential is closely related to selection rules for harmonic generation in two-color fields (see \cite{cecch}, and references therein). If the two fields with frequencies $\omega_1$ and $\omega_2=n\omega_1$ are equally polarized instead of oppositely, the KH potentials have  ($n-1$)-fold symmetry. Hence, to obtain ``trichotomy'' $\omega_2=4\omega_1$ has to be chosen in this case. If the two frequencies are not commensurable the free electron trajectories are not closed, and the two-color, time-average KH potential is ill-defined. However, we observed stabilization also in this case although no distinct probability density peaks were obtained.

\subsection{Snapshots of the probability density}
\begin{figure}
\includegraphics[scale=0.5]{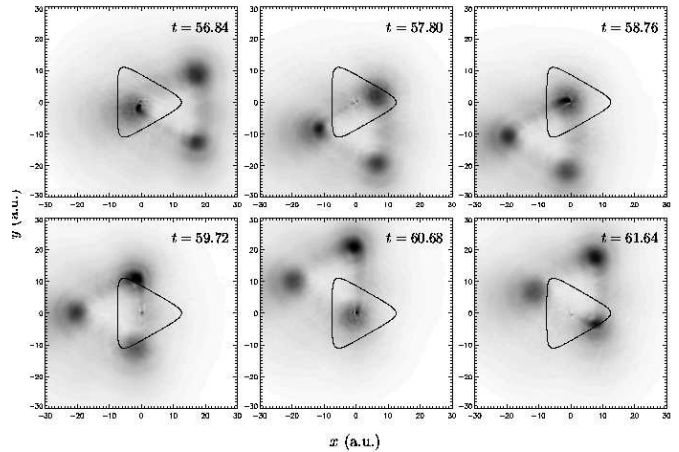}
\caption{\label{snapshots:10:10:trichotomy} Series of contour plots
of the probability density in the $xy$-plane at $z=0$  during the 9th (low frequency) cycle of the laser pulse for $\omega_1=1$ [(2,8,2)-pulse],
$\omega_2=2$ [(4,16,4)-pulse], and $\Edach_1=\Edach_2=10$ ``Trichotomy'' and ``hula-hoop''-dynamics is observed, i.e., three peaks of probability density, located in the
corners of an equilateral triangle whose center moves along the
black triangle representing the orbit of a free electron in the same
laser field. The color-coding of the probability density is linear.}
\end{figure}

In Fig.~\ref{snapshots:10:10:trichotomy} snapshots of the trichotomous ``hula-hoop''-dynamics for $\omega_1=1$, $\omega_2=2$ during one laser cycle (with respect to the $\omega_1=1$-field)  is presented. The two laser intensities were chosen equal so that the free-electron trajectory is a triangle with rounded edges. This triangle is drawn in black in each of the plots of Fig.~\ref{snapshots:10:10:trichotomy}. As expected, the center of the trichotomous probability density (aligned along the time-averaged KH potential) moves clockwise along the black triangle. Note that the stabilized structure moves as a whole and does not rotate about its center. In the comoving frame, i.e.,  the KH frame,  the ion swirls along the triangle, forming another probability density maximum.

\begin{figure}
\includegraphics[scale=0.5]{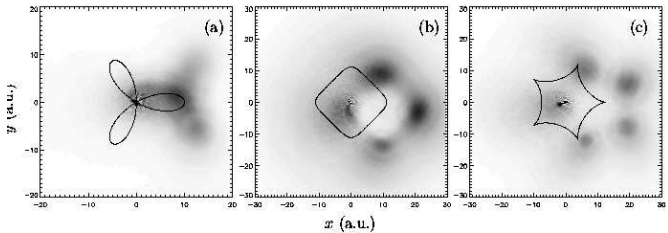}
\caption{\label{triquatropenta} Probability densities in the $xy$ plane at $z=0$ and $t=56.84$ during the constant part of a (2,8,2)-pulse (with respect to $\omega_1=1$). The second frequency $\omega_2$ was (from left to right) 2,3, and 4. The laser intensities were chosen such that (a) $\alphahat^{(1)}=\alphahat^{(2)}=5$, (b) $\Ehat^{(1)}=\Ehat^{(2)}=10$, and (c) $\Ahat^{(1)}=\Ahat^{(2)}=10$. The overlayed black curves are the corresponding free electron trajectories along which the center of the stabilized structure moves.  
}
\end{figure}

In Fig.~\ref{triquatropenta} three other examples for stabilized probability density are presented. In panel (a), where the excursion amplitudes for both fields were chosen equal, no three distinct peaks are visible because $\alphahat=5$ is too small. In (b) a clear example of equal intensity quatrochotomy (i.e., $\omega_2=3$) is shown and in (c) a case of pentachotomy ($\omega_2=4$) where both fields had equal vector potential amplitudes. The density of the $n+1$ peaks varies during the course of the laser pulse. This implies that several Floquet states are involved.

\subsection{Survival probability vs.\ time}
\begin{figure}
\includegraphics[scale=0.75]{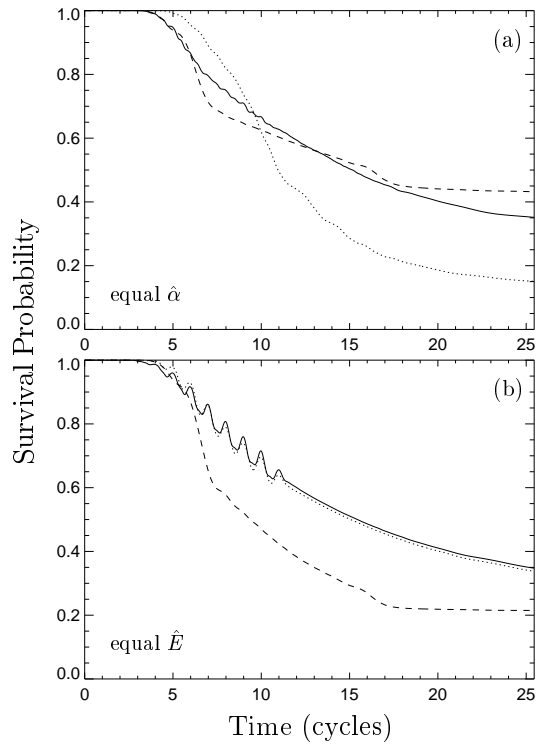}
\caption{\label{comp_oneandtwocolor} The probability to survive the laser field non-ionized vs.\ time in laser cycles of the lower frequency field with $\omega_1=1$. The full curves are the two-color results for $\omega_2=2$, the dotted curves are the results for the low-frequency field alone, the dashed curves are the results for the $\omega_2=2$-field alone. In panel (a) the equal excursion amplitude case $\alphahat=2$ is shown while in (b) the equal intensity case $\Ehat=6$ is presented. (2,8,2)- and (4,16,4)-pulses were used for the $\omega_1$- and $\omega_2$-field, respectively. 
}
\end{figure}

So far the two-color survival probability was not addressed at all. If, for instance, $\Ehat^{(2)}$ is chosen so that the $\omega_2=2$-field alone yields optimal stabilization, and a $\omega_1=1$-field with $\Ehat^{(1)}$ in the ``death valley'' is added---what is the survival probability? In the worst case ionization would be dominated by the ``death valley''-field, making the two-color stabilization not very attractive. Fortunately, it is not the field which would lead to highest ionization, if applied alone, that dominates two-color dynamic stabilization.  
In Fig.~\ref{comp_oneandtwocolor} the temporal evolution of the survival probability in the two-color case is compared with the two corresponding single-color results. From panel (a) it is clearly seen that if the two laser intensities are tuned to yield equal excursion amplitudes $\alphahat$ the two-color survival probability is closer to the single-color high-frequency result with relatively low ionization probability. The low-frequency field alone has a higher ionization probability. In the equal intensity plot (b) it is the other way round: the two-color result follows closely the low-frequency curve whereas the high-frequency field alone yields higher ionization because $\Ehat^{(2)}$ is still close to the ``death valley.'' Since in plot (a) all curves cross each other it is apparent that the final survival probability is sensitive to the pulse length. However, it should be stressed that, fortunately,  the two-color survival probability is {\em not} determined by the frequency which yields highest ionization. This means that the two-color ``probability density shaping,'' as discussed in the previous subsection, can be performed without drastic reduction of the survival probability.  

\begin{figure}
\includegraphics[scale=0.5]{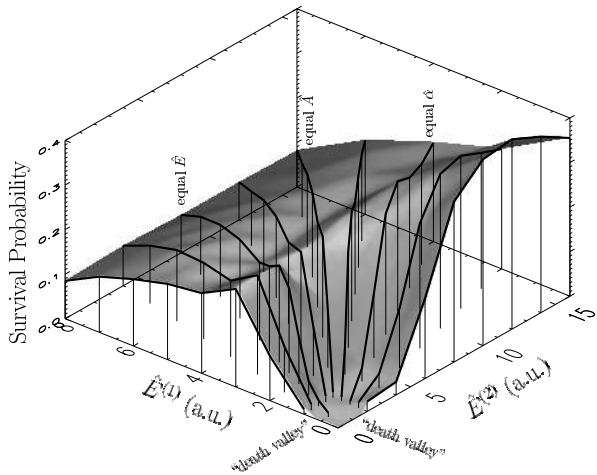}
\caption{\label{survprobvsE1E2} The survival probability after a (2,8,2)- and (4,16,4)-pulse of frequency $\omega_1=1$ and $\omega_2=2$, respectively, vs.\ the peak field strengths $\Ehat^{(1)}$ and $\Ehat^{(2)}$. The data points indicated by vertical lines were calculated, the shaded surface is an interpolation between them. Increasing survival probability for $\Ehat^{(1)},\Ehat^{(2)}\to 0$ was suppressed for the sake of better visibility of the results in the stabilization regime. See text for discussion.
}
\end{figure}

In Fig.~\ref{survprobvsE1E2} the survival probabilities at the end of (2,8,2)- and (4,16,4)-pulses for $\omega_1=1$ and $\omega_2=2$, respectively, are plotted vs.\ the field amplitudes $\Ehat^{(1)}$ and $\Ehat^{(2)}$. Because of the probability density representing free but slow electrons it takes a long time until the probability density integrated over the sphere of radius $R_b=60$ assumes a constant value. Hence, to obtain  Fig.~\ref{survprobvsE1E2} we have chosen to measure the survival probability as the maximum probability to find the electron in a smaller sphere of radius $R_s=7.5$ within a time period of 85\,a.u.\ after the pulse. The maximum probability has to be taken because of oscillations due to excitation of the 2s and the 3s state. Note that this method for measuring the survival probability slightly overestimates ionization because probability density which represents slow recombining electrons, slowly spirals back to the ion and does not enter the small sphere within the observation time. However, here we are not interested in absolute numbers (which depend sensitively on the pulse duration and shape anyway). From Fig.~\ref{survprobvsE1E2} one infers that the highest survival probability is obtained, as expected, in the single-color high-frequency case where $\Ehat^{(1)}=0$. In general, the overall behavior is an increasing stabilization probability up to a maximum value, followed by increasing ionization if the laser intensities are increased further. This is different from high-frequency Floquet theory where the ionization rate monotonically decreases with increasing intensity because the ``shake-off'' during the pulse turn-ons and -offs is not taken into account.

In our results, the field strength regime where $\Ehat^{(1)}/\omega_1,\Ehat^{(2)}/\omega_2<1$ in Fig.~\ref{survprobvsE1E2} may be identified as the ``death valley'' of high ionization probability. Of course, the survival probability increases towards unity as $\Ehat^{(1)},\Ehat^{(2)}\to 0$. However, this has been  suppressed in the plot because it blocked the visibility of the results in the stabilization regime in which we are interested. Given a certain $\Ehat^{(1)}$, adding $\Ehat^{(2)}$ is, from a stabilization point of view, almost always beneficial, provided that $\Ehat^{(2)}$ is not too strong so that shake-off dominates. If, on the other hand, we choose a certain $\Ehat^{(2)}$ and add a $\omega_1$-field the survival probability may be enhanced or diminished, depending on whether $\Ehat^{(2)}$ lies in its single-color stabilization regime or near its single-color ``death valley.'' The case of a rosette-like stabilized probability density where the excursion amplitudes $\alphahat$ are equal leads only to a relatively small reduction in the survival probability  with respect to the optimally stabilized $\omega_2$-field alone (around $\Ehat^{(2)}=12$ in Fig.~\ref{survprobvsE1E2}). In contrast, the equal intensity result remains close to the low-frequency single-color result where $\Ehat^{(2)}=0$. Note that maximum stabilization is achieved for modest values of $\alphahat=\Ehat/\omega^2$ where the probability density splitting, i.e., trichotomy, quatrochotomy, etc., is not yet developed.

\subsection{ATI in circular two-color stabilization}
Finally, let us briefly discuss the two-color photoelectron spectra. The single color case has been analyzed in subsection \ref{ati_singlecolor}. 

\begin{figure}
\includegraphics[scale=0.5]{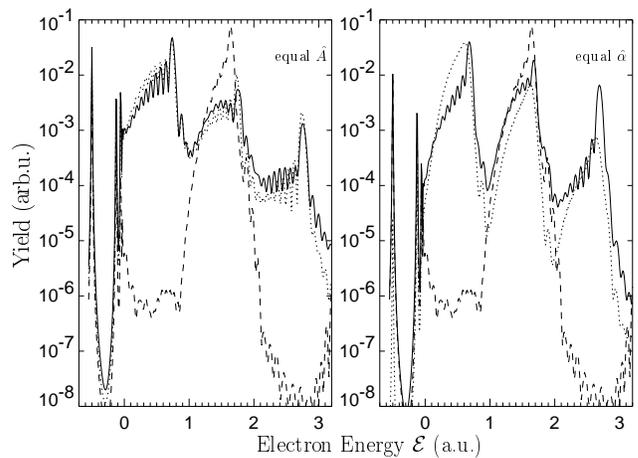}
\caption{\label{compati} Comparison of two-color photoelectron spectra (full curve) with the corresponding single-color results where only the $\omega_1=1$-field (dotted) and only the $\omega_2=2$-field (broken) was present, respectively. The same pulse shapes as in Fig.~\ref{atispectra} were used. The two-color result in the left panel has been obtained with $\Adach=\Ehat^{(1)}/\omega_1=\Ehat^{(2)}/\omega_2=2$. In the right panel the excursion amplitudes were equal: $\alphahat=\Ehat^{(1)}/\omega_1^2=\Ehat^{(2)}/\omega_2^2=1$.
}
\end{figure}

In Fig.~\ref{compati} the two-color spectra are compared with the corresponding single-color results where either the $\omega_1=1$-field  or the $\omega_2=2$-field was present. The laser intensities were chosen not deep in the stabilization regime  because at such high laser intensities the clear ATI-structure gets corrupted by peak broadening and modulations  (cf.\ Fig.~\ref{atispectra}). 

The ATI peaks in the two-color spectra are located at higher energies than those of the corresponding single-color results. This is due to the fact that the ground state in the time-averaged KH potential experiences a stronger up-shift as compared to the single-color KH potentials. It would be interesting to analyze the ATI peak positions in the two-color case in the same way  as it was done in Sec.\ \ref{ati_singlecolor} for single-color fields. However, to our knowledge two-color Floquet energies for high-frequency, circular polarization are not yet published.  

As far as the peak strengths are concerned the two-color spectrum is, of course,  not simply the sum of the two single-color spectra: the first ATI peak in the single-color $\omega_2$-spectrum is stronger than both, the second peak of the two-color and the single-color $\omega_1$-field. This is compatible with Fig.~\ref{survprobvsE1E2} from which one infers that adding to the $\omega_2$-field (whose intensity lies close to the ``death valley'') the lower frequency $\omega_1$-field  suppresses ionization. In the equal-$\Ahat$-result the two-color spectrum is close to the single-color, low-frequency spectrum. Instead, the two-color spectrum for equal $\alphahat$ is qualitatively different from both  two single-color spectra. Not only modulations appear in the two-color result which are absent in the $\omega_1$-spectrum but also the height of the peaks differ significantly from each other.

\section{Conclusion} \label{concl}
We have studied the dynamic stabilization of atomic hydrogen against ionization in single- and two-color, circularly polarized laser pulses by solving numerically the three-dimensional, time-dependent Schr\"odinger equation. In the single-color case we confirmed the ``hula-hoop''-like dynamics for sufficiently high excursions $\alphahat$. Ionization rates were compared with those from (high-frequency) Floquet theory, and good agreement was found. The positions of above-threshold ionization (ATI) peaks in our numerical photoelectron spectra were analyzed. It was found that the ATI peaks move with increasing laser intensity towards higher energies, in contrast to what happens in intense, low-frequency fields. This behavior can be explained by the laser field-induced up-shift of the initially populated, field-free  ground state.

For two-color laser fields of
opposite circular polarization we presented stabilized probability density structures. For laser frequencies $\omega_1$ and  $\omega_2=n\omega_1$, $n=2,3,\ldots$ and sufficiently large excursion amplitudes $(n+1)$ distinct probability density peaks were observed, in accordance to what one expects by virtue of the two-color, time-averaged Kramers-Henneberger potentials. As the generalization  of the well-known ``dichotomy'' in linearly polarized laser fields, we called this multiple probability density peak-splitting  ``trichotomy,'' ``quatrochotomy,'' ``pentachotomy'' etc. 

The survival probability after two-color pulses of frequencies $\omega_1=1$ and $\omega_2=2$ was calculated as a function of the two peak field strengths. As expected, highest stabilization is achieved with the high-frequency field alone for a certain, optimal laser intensity. However, adding the second, low-frequency field, even if its intensity falls into the ``death valley,'' fortunately does {\em not} drastically reduce stabilization, as one might expect from a naive superposition point-of-view.

\begin{acknowledgments}
The authors thank Dr.\ R.\ Potvliege for providing electronically the Floquet rates and energy shifts to which comparison was made in Sec.~\ref{singlecolor}. The permission to run our codes  on the Linux cluster at PC$^2$ in Paderborn, Germany, is gratefully acknowledged. This work was supported in part by the INFM through the Advanced Research Project CLUSTERS. 
\end{acknowledgments}



\begin{thebibliography}{}
\bibitem{pont89} M.\ Pont, \pra {\bf 40}, 5659 (1989).
\bibitem{pontgav90} M.\ Pont and M.\ Gavrila, \prl {\bf 65}, 2362 (1990).
\bibitem{pontwalet} M.\ Pont, N.\ R.\ Walet, and M.\ Gavrila, \pra {\bf 41}, 477 (1990).
\bibitem{gavrevI} Mihai Gavrila, in: {\em Atoms in Intense Laser Fields}, ed.\ by M.\ Gavrila (Academic Press, San Diego, 1992).
\bibitem{eberlykul} J.\ H.\ Eberly and K.\ C.\ Kulander, Science {\bf 262}, 1229 (1993).
\bibitem{mittlemanbook} Marvin H.\ Mittleman, {\em Introduction to the Theory of Laser-Atom Interactions}, 2nd ed.\ (Plenum Press, New York, 1993).
\bibitem{faisal} F.\ H.\ M.\ Faisal, L.\ Dimou, H.-J.\ Stiemke, and M.\ Nurhuda, J.\ Nonl.\ Opt.\ Phys.\ Mat.\ {\bf 4}, 701 (1995).
\bibitem{gavrilarevII} M.\ Gavrila, in: {\em Multiphoton Processes}, ed.\ by L.\ F.\ DiMauro {\em et al.} AIP Conf.\ Proc.\ No.\ 525 (AIP, Melville, NY, 2000). 
\bibitem{kulander} Kenneth C.\ Kulander, Kenneth J.\ Schafer, and Jeffrey L.\ Krause, \prl {\bf 66}, 2601 (1991).
\bibitem{dondera} M.\ Dondera, H.\ G.\ Muller, and M.\ Gavrila, \pra {\bf 65}, 031405 (2002). 
\bibitem{choi} Dae-Il Choi and Will Chism, \pra {\bf 66}, 025401 (2002).
\bibitem{druten} N.\ J.\ van Druten, R.\ C.\ Constantinescu, L.\ D.\ Noordam, and H.\ G.\ Muller, \pra {\bf 55}, 622 (1997), and references therein. 
\bibitem{doerr} Martin D\"orr, R.\ M.\ Potvliege, Daniel Proulx, and Robin Shakeshaft, \pra {\bf 43}, 3729 (1991).
\bibitem{su} Q.\ Su, J.\ H.\ Eberly, and J.\ Javanainen, \pra {\bf 64}, 862 (1990).
\bibitem{volkova} E.\ A.\ Volkova, A.\ M.\ Popov, and O.\ V.\ Smirnova, JETP {\bf 79}, 736 (1994) [Zh.\ Eksp.\ Teor.\ Fiz.\ {\bf 106}, 1360 (1994)].
\bibitem{patelI} A.\ Patel, M.\ Protopapas, D.\ G.\ Lappas, and P.\ L.\ Knight, \pra{\bf 58}, R2652 (1998).
\bibitem{patelII} A.\ Patel, N.\ J.\ Kylstra, and P.\ L.\ Knight, Opt.\ Express {\bf 4}, 496 (1999); J.\ Phys.\ B {\bf 32}, 5759 (1999).
\bibitem{kwon} Duck-Hee Kwon, Yong-Jin Chun, Hai-Woong Lee, and Yongjoo Rhee, \pra {\bf 65}, 055401 (2002).
\bibitem{ryab} M.\ Yu. Ryabikin and A.\ M.\ Sergeev, Laser Physics {\bf 12}, 757 (2002).
\bibitem{kylstra} N.\ J.\ Kylstra, R.\ A.\ Worthington, A.\ Patel, P.\ L.\ Knight, J.\ R.\ V\'azquez de Aldana, and L.\ Roso, \prl {\bf 85}, 1835 (2000).
\bibitem{bauer} D.\ Bauer and F.\ Ceccherini, \pra {\bf 60}, 2301 (1999).
\bibitem{cheng} Taiwang Cheng, Jie Liu, and Shigang Chen, \pra {\bf 59}, 1451 (1999).
\bibitem{potvl} R.\ M.\ Potvliege, \pra {\bf 60}, 1311 (1999).
\bibitem{bauerhabil} D.\ Bauer, {\em Intense laser field-induced multiphoton processes in single- and many-electron systems} (Habilitation thesis, TU Darmstadt, 2002); submitted for publication.
\bibitem{mullermethod} H.\ G.\ Muller, Laser Physics {\bf 9}, 138 (1999).
\bibitem{windowop} K.\ J.\ Schafer and K.\ C.\ Kulander, \pra {\bf 42}, 5794 (1990).
\bibitem{cecch} F.\ Ceccherini, D.\ Bauer, and F.\ Cornolti, J.\ Phys.\ B {\bf 34}, 5017 (2001).
\end{thebibliography}
\end{document}